\documentclass[preprint]{jpsj3}
\usepackage{txfonts}
\usepackage{graphicx}
\usepackage{color}

\def\tw {\textwidth}

\title{Unusual Superconducting Proximity Effect\\ in Magnetically Doped Topological Josephson Junctions}

\author{Rikizo Yano$^{1, 2}$\thanks{yano-rikizo@nagoya-u.jp}, Masao Koyanagi$^1$, Hiromi Kashiwaya$^1$, Kohei Tsumura$^{1, 2}$, \\
Hishiro T. Hirose$^3$, Yasuhiro Asano$^4$, Takao Sasagawa$^3$, and Satoshi Kashiwaya$^{1, 2}$\thanks{s.kashiwaya@ap.pse.nagoya-u.ac.jp}}

\inst{$^1$National Institute of Advanced Industrial Science and Technology (AIST), Tsukuba 305-8568, Japan \\
$^2$Department of Applied Physics, Nagoya University, Nagoya 464-8603, Japan \\
$^3$Laboratory for Materials and Structures, Tokyo Institute of Technology, Yokohama 226-8503, Japan \\
$^4$Center of Topological Science and Technology, Hokkaido University,  Sapporo 060-8628, Japan \\} 

\abst{The transport properties of a topological Josephson junction fabricated from a magnetically doped topological insulator (TI) were investigated. The conductance spectra of the Nb/Fe-Bi$_2$Te$_2$Se/Nb junction below 1 K showed an unusual trident-shaped zero-bias conductance peak with a tiny peak width of $\sim$ 6 $\mu$V. The central peak of the trident peak presents the dc-Josephson current, and the side peaks may reflect an induced unconventional Cooper pairing. Additionally, the critical currents followed inverse to temperature, which may also reflect the presence of an unconventional proximity effect. Furthermore, microwave irradiation derived a drastic change in the conductance spectra from the peak structure into oscillatory ones, a hallmark of the ac-Josephson supercurrent. The current-phase relation of the ac-Josephson effect under high power radiofrequency-irradiation was found to be 4$\pi$-periodic. The results suggest that the junction based on magnetically doped 3D TIs may realize an unconventional Cooper pairing, thus enabling access to the basic physics of Majorana bound states and unconventional superconductivity.}

\begin{document}
\maketitle

\section{Introduction}

Probing S/I/N/I/S junction (S: superconductor, I: Insulator, N: normal metal) is one of the powerful tools to investigate the existence of Andreev bound states (ABS) and current-phase relations in unconventional superconductor candidates.~\cite{ref1, ref2} When the barrier height of the insulator is high and the ratio of coherence length and an electrode gap ($2\xi_N/ L$) is less than unity, i.e., a tunneling (S/I/N) condition, the conductance spectra of the junction reflect density of states (DOS) of quasiparticles,~\cite{ref3} while, Josephson supercurrent emerges in a low barrier condition and $2\xi_N/ L > 1$ (an S/N/S condition).~\cite{ref4} Practically, a junction with a moderate barrier height and $L$ shows both characteristics.

As hallmarks of unconventional superconducting (SC) states, two types of signatures have been explored. In conductance measurements of S/I/N (S/N) junctions, an unconventional superconductor is expected to show an unique zero-bias conductance peak (ZBCP), reflecting the symmetry of SC gap functions and surface Andreev bound states (ABSs).~\cite{ref5, ref6, ref7, ref8} When the ABS has a special zero energy states called Majorana bound state (MBS), the appearance of 4$\pi$-periodic Josephson supercurrent is expected in the topological Josephson (S/N/S) junction.~\cite{ref9} Recently, such signature behaviors are observed in nanowires with strong spin-orbit interaction (SOI)~\cite{ref10} and spin Hall insulator systems.~\cite{ref12} Those systems emulate $p$-wave superconductors. As a result, Majorana fermions, whose anti-particles are identical to the fermions themselves,~\cite{ref13} have been expected to emerge at the edges of the $p$-wave superconductors.~\cite{ref9, ref14}

In the context of searching for unconventional SC states that have the MBS, topological insulators (TIs) are promising material systems for the following three reasons. First, the emergence of a proximity-induced chiral  $p$-wave-like SC state preserving time-reversal symmetry (TRS), and the Majorana fermions at the two-dimensional (2D) interface of TI/S were predicted theoretically.~\cite{ref15} Indeed, several studies report observing a ZBCP~\cite{ref16, ref17, ref18, ref19, ref20} and a sign-changing order parameter implying the chiral  $p$-wave component.~\cite{ref21} Second, the 2D surface of 3D-TI is favorable for tuning experimental set-ups and configurations, for future braiding operations using complicated electrodes on chips.~\cite{ref22, ref23, ref24, ref25, ref26} Finally, odd-frequency Cooper pairing may exist at the boundary of the TI/S. In the symmetry classification of pairing, Cooper pairs can be categorized into four types depending on their symmetries with respect to Matsubara frequency, spin, and orbital.~\cite{ref27} For example, $s$- and  $p$-wave pairs in bulk are written as even-frequency spin-singlet even-parity (ESE) and even-frequency spin-triplet odd-parity (ETO), respectively, corresponding to their parity of the above three symmetries. Odd-frequency Cooper pairs (odd-singlet-odd (OSO) and odd-triplet-even (OTE)) are closely related to the physics of ABS and Majorana fermions.~\cite{ref27, ref28} Due to the strong SOI of TI, ESE pairing is supposed to be suppressed, and other unconventional pairs emerge at the TI/S interface.

Unconventional superconductors may also be realized using magnetic moments. Induced Zeeman splitting leads to equal-spin triplet ($\uparrow  \uparrow$) components instead of the conventional $s$-wave pairing. In the ultimate case, a dominant OTE pairing was theoretically deduced for an S/half-metal/S junction,~\cite{ref29} and this was confirmed experimentally.~\cite{ref30} However, an eccentric phenomenon derived from the dominant OTE component is yet to be reported.

The combination of magnetic moments and spin-momentum locking in TIs may lead to rather robust unconventional Cooper pairs. According to a theoretical study, the majority component depends on the orientation of magnetic moments; e.g., in-plane magnetic moments lead to OTE pairing.~\cite{ref31} Besides, the emergent proximity effects are expected to depend on the distances from SC electrodes as described by Shen et al.~\cite{ref32} (Fig.\ref{fig1cap}(e)); therefore, Josephson junctions with an appropriated gap distance $L$, should show a unique, unconventional SC nature. However, a Josephson junction consisting of magnetic 3D-TI has not yet been explored. 

In this paper, we report the transport properties of an S/Ferromagnetic-TI(FTI)/S junction fabricated from a Fe-doped Bi$_2$Te$_2$Se having high bulk insulating properties. Two types of spectral features coexist because of the moderate gap length of electrodes: i.e., the conductance peak coming from a Josephson effect and another origin. The spectra showed an unusual trident-shaped ZBCP and 4$\pi$-periodic Josephson supercurrent, which suggests the existence of unconventional Cooper pairs caused by proximity effect in the magnetic TI junction.

\section{Experiment}

The single crystals of Bi$_{1.99}$Fe$_{0.01}$Te$_2$Se (Fe-BTS) were grown by a modified Bridgman method.~\cite{ref33} Ferromagnetism of the crystals was checked using a commercial SQUID magnetometer (Quantum Design: MPMS-XL). {S/FTI/S junctions, as described in Fig.\ref{fig1cap}(a), were fabricated using exfoliated Fe-BTS crystals, followed by Nb-sputtering. The crystals were exfoliated onto a SiO$_{2}$/Si substrate. Larger SC-electrodes were made by sputtering with Nb (50 nm)/Ti (10 nm). The narrower SC-parts of Ti (5 nm)/Nb (90 nm)/Ti (5 nm) were deposited onto the crystal via electron beam lithography and lift-off technique. As a result, S/FTI/S Josephson junctions with approximately 100 nm length gap and 300 nm width, were obtained, as shown in Fig.\ref{fig1cap}(b). The junction between terminals 1 and 3 was mainly focused in this paper.The transport properties of the junctions were measured by a standard lock-in technique with a modulation frequency of 761.1 Hz. Microwave (radio frequency: rf) was irradiated via a loop antenna with a diameter of $\sim$ 10 mm, and a magnetic shield (the maximum magnetic field is limited to 50 G) was used to reduce external magnetic noises.

\section{Results}

\begin{figure}[h]
\centering
\includegraphics[width=0.6\tw]{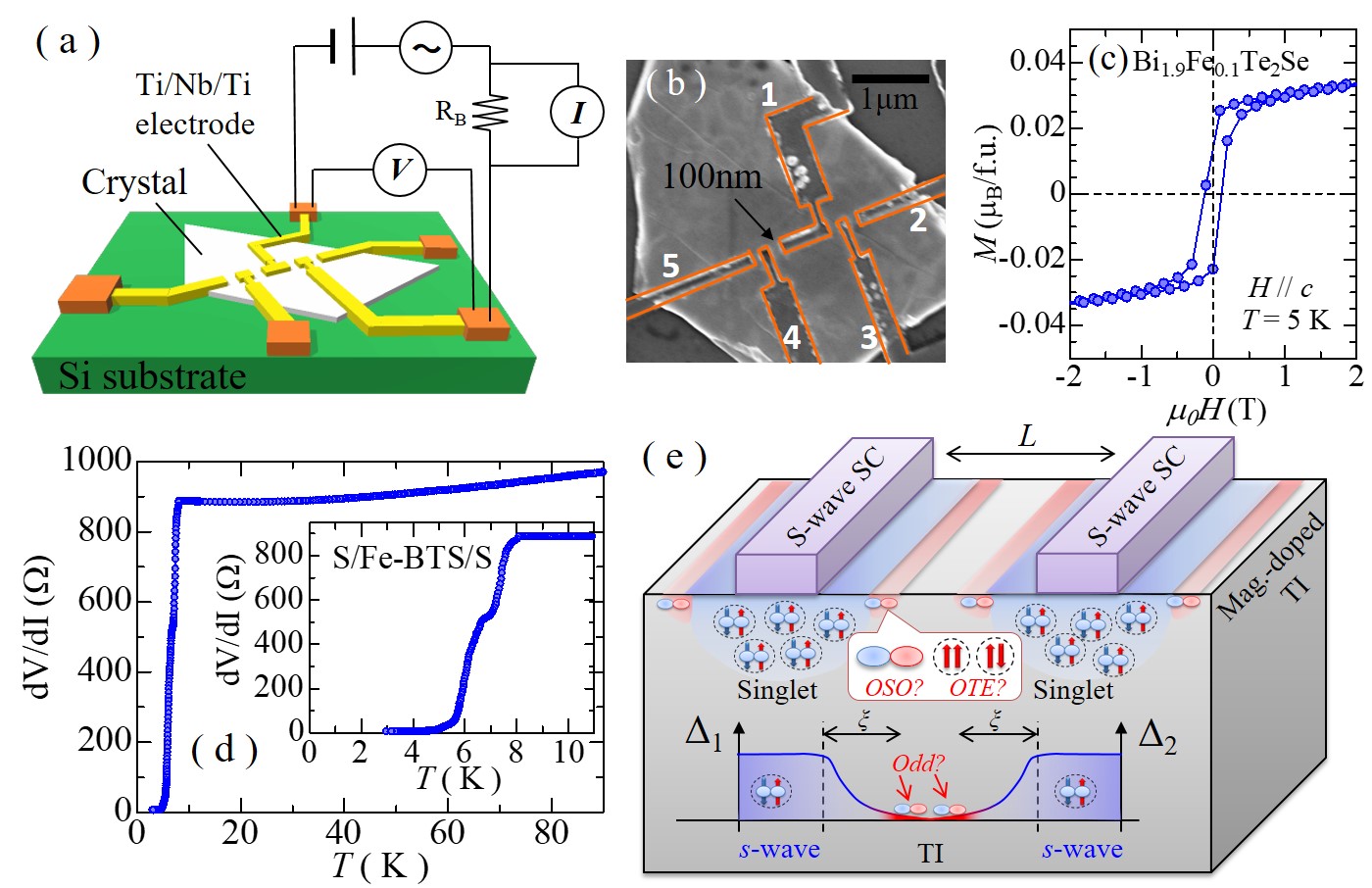}
\caption{\label{fig1cap}  (Color online) (a) The schematic device and circuit layout. (b) Top view of fabricated S/Fe-BTS/S junctions. Orange lines highlight electrodes. (c) $c$-axis magnetization of a bulk Fe-BTS singe crystal at 5 K. (d) Temperature dependence of zero-bias resistance on the S/FTI/S junction (1-3 terminal). Low temperature region is highlighted in inset. (e) The schematic model of the proximity effect and estimated pairing on the S/FTI/S junction.}
\end{figure}

The resulting crystals of Fe-BTS showed $n$-type conduction with high resistivity ($\sim$ 0.20 m$\Omega$m at 300 K), a carrier density $n$ of  5.6  $\times$ 10$^{24}$ m$^{-3}$, and total bulk mobility of $\mu_{\mathrm{b}}$ $\sim$ 5600 mm$^2$/Vs. These values were compatible with a typical large bulk insulating TI.~\cite{ref34, ref35} Figure \ref{fig1cap}(c) shows the substantial ferromagnetic magnetization of the crystals similar to that of Fe-doped Bi$_2$Se$_3$,~\cite{ref33} which establishes magnetically doped TI with high resistivity. The coercive field was $\sim$0.15 T (8.7 $\mu$eV). The easy-axis of the bulk crystal was along the $c$-axis, similar to the Fe-doped Bi$_2$Se$_3$.~\cite{ref33}

The band structure of the bulk crystal was evaluated. Non-doped BTS has the Dirac cone with band gap ($E_{\mathrm{g}} = 215$ meV)~\cite{ref36}, and broken TRS by Fe-doping should lead to a gap opening at the Dirac point.~\cite{ref33} Such gap-opening at the Dirac point and comparable $E_{\mathrm{g}}$ were confirmed in our samples by a corroborating angle-resolved photoemission spectroscopy study ($E_{\mathrm{mag}} \sim 20-50$ meV). The Fermi level $E_{\mathrm{F}}$ was located above the gaped Dirac point.

Important parameters that determine junction nature were calculated using the transport properties of bulk crystals. The mean free path $\ell_{\mathrm{3D}}$ of $\sim$ 8.5 nm at 2 K. Some crystals showed 2D weak anti-localization nature in magnetoresistance measurements. The phase-coherent length was obtained as $\ell_\phi \sim$ 37 nm by its fitting result. Because Shubnikov-de Haas oscillations were not observed up to 9 T, we did not obtain the exact coherence length $\xi_{\mathrm{N}}$, which determines the propagation distance of the proximity effects. Instead, we estimated the general coherence length $\xi_{\mathrm{N,g}}$ using the above experimental parameters and reported values.~\cite{ref34, ref37} Considering Refs.~\citen{ref38} and~\citen{ref39}, the $\xi_{\mathrm{N,g}}$ can be calculated by using the following formula:
\[\displaystyle \xi _{\mathrm{N,g}} = (\frac{\hbar ^3 \mu ^2 \nu _\mathrm{F}}{2\pi k_\mathrm{B} T e^2 \ell_{\mathrm{3D}} })^{\frac{1}{2}}\! \times \! (3\pi ^2 n)^{\frac{1}{3}} \!\times \!(1\! +\! \frac{2\pi k_\mathrm{B} T \ell_{\mathrm{3D}}}{\hbar \nu_\mathrm{F} })^{-\frac{1}{2}},\]
where $k_\mathrm{B}, T, e, \hbar$, and $\nu_\mathrm{F}$ are the Boltzmann constant, temperature, the elementary charge, the reduced Planck constant, and the Fermi velocity, respectively.  This expression covers both dirty and clean limits. As a result, the $\xi_{\mathrm{N,g}}$ of Fe-BTS was less than 40 nm. 

The fabricated S/FTI/S junctions with $L =$ 100 nm exhibited metallic temperature dependence (Fig.~\ref{fig1cap}(d)). The resistance value was reasonably consistent with the expected resistance from the resistivity for bulk crystals within an order of magnitude. A clear two-step transition ($T_{\mathrm{c1}} =$ 7.4 K, and $T_{\mathrm{c2}} =$ 6.0 K; both were lower than the critical temperature of bulk Nb, $T_{\mathrm{c}}$(Nb)$ =$ 9.2 K~\cite{ref40}), was widely observed in the Nb-based junctions. The resistance below $T_{\mathrm{c2}}$ was not zero, but $\sim$ 1.0 $\Omega$, even at the lowest temperature of $T =$ 0.55 K. Note that we repeatedly confirmed limited contribution of measurement condition, such as noise, to the finite resistance and concluded that the resistance represents the intrinsic nature of the junction. As discussed later, this resistance may reflect the nature of a phase-slip Josephson junction.


\begin{figure}[h]
\centering\includegraphics[width=0.6\tw]{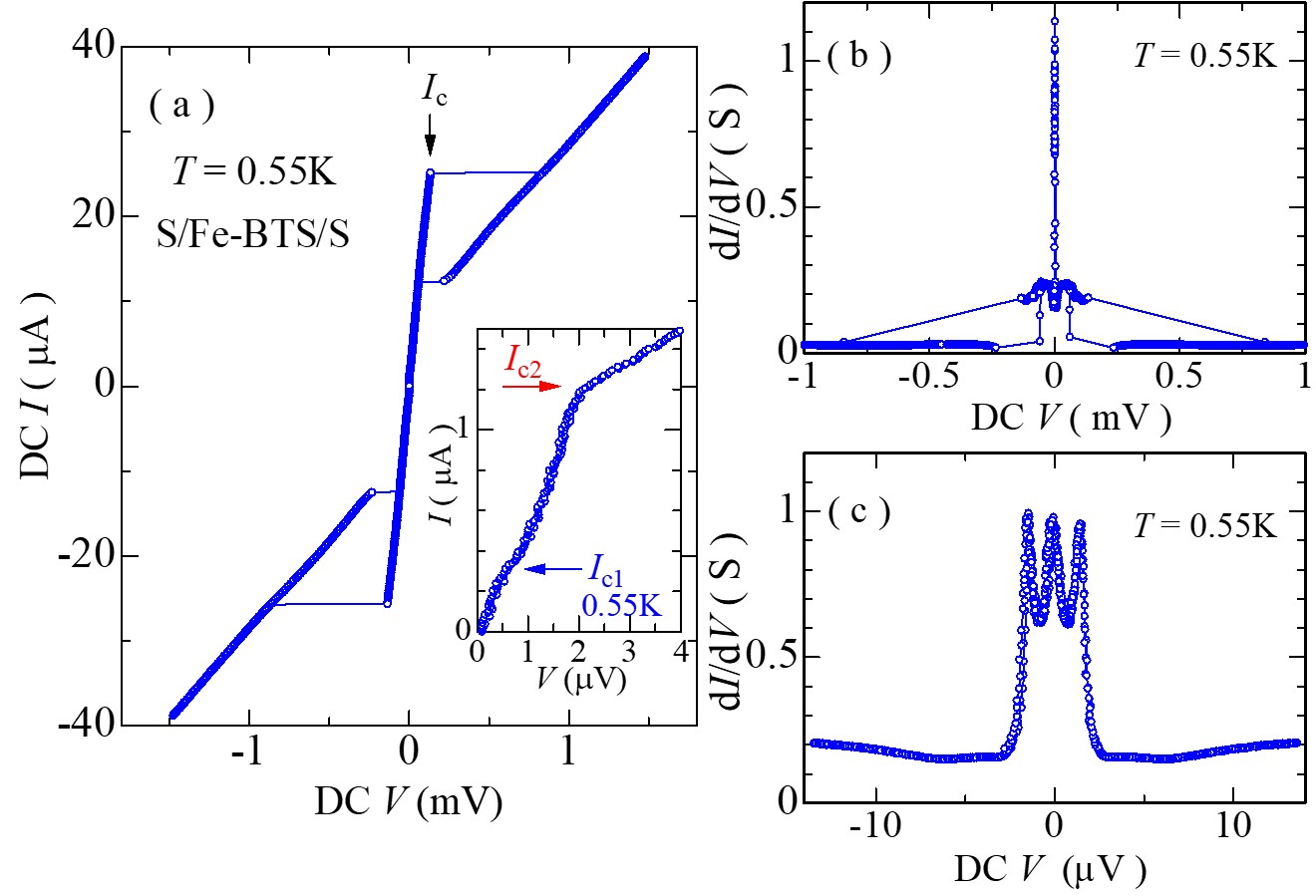}
\caption{\label{fig2}   (Color online) (a) An $I$-$V$ curve of S/Fe-BTS/S junction at 0.55 K. Arrow indicates the critical current of $I_{\mathrm{c}}$. Insets indicates a highlighted induced structure at several temperatures. Blue and red arrow indicate characteristic currents for $I_{\mathrm{c1}}$ and $I_{\mathrm{c2}}$, respectively. (b) dc-Voltage dependence of deferential resistance $dI/dV$-$V$ of the S/FTI/S junction at 0.55 K. Low-bias part of the $dI/dV$-$V$ curve showing the induced peak is highlighted in (c). }
\end{figure}

At the lowest temperature of 0.55 K, an $I$-$V$ curve showed a remarkable hysteresis of the Josephson effect (Fig.~\ref{fig2}(a)). This hysteretic jump was also found in the $dI/dV$ measurements (Fig.~\ref{fig2}(b)). The critical current $I_{\mathrm{c}}$ was 24 $\mu \mathrm{A}$ and the product of $I_{\mathrm{c1}}$ and resistance of a normal state ($I_{\mathrm{c}}R_{\mathrm{N}}$), which reflects the energy scale of pair potential $\Delta$(T), was calculated to be 1.02 meV at 0.55 K. This value is comparable with the SC gap of Nb ($\Delta _{\mathrm{Nb}} =$ 1.51 meV~\cite{ref40}). The hysteretic structure vanished above 8 K. The temperature dependence of the product $I_{\mathrm{c}}R_{\mathrm{N}}$ obeyed the conventional Ambegaokar-Baratoff theory.~\cite{ref41} Those behaviors are consistent with the conventional Josephson effect, and as discussed later, those represented nature of Nb-electrodes, not represented proximity effect.

Anomalous proximity effects were discovered at a small energy scale. The inset of Fig.~\ref{fig2}(a) shows two bend structures as characterized by blue ($I_{\mathrm{c1}}$) and red ($I_{\mathrm{c2}}$) arrows in a small bias region of the $I$-$V$ characteristics. Similar bending is widely observed in SC weak links  (S/N/S junctions). We performed differential conductance measurements to highlight this small energy structure. While the narrow structure seems like an extraordinary sharp ZBCP in the wide-ranging scale shown in Fig.~\ref{fig2}(b), the peak had a trident shape and shallow dips just outside of the peak ($\sim \pm$7 $\mu $V) in the highlighted scale (Fig.~\ref{fig2}(c)). The peak width was $\sim$6 $\mu $V, which is quite small, less than the thermal energy ($\sim$47.4 $\mu $eV at 0.55 K). We confirmed repeatability of this trident shaped structure using other Nb/Fe-BTS/Nb junctions, and almost the same structure appeared even in a Josephson junction fabricated on further surface dominant magnetic TI ~\cite{newref41}. This fact suggests that the unique peaks should be originated from the dominant surface states.

\begin{figure}[h]
\centering\includegraphics[width=0.6\tw]{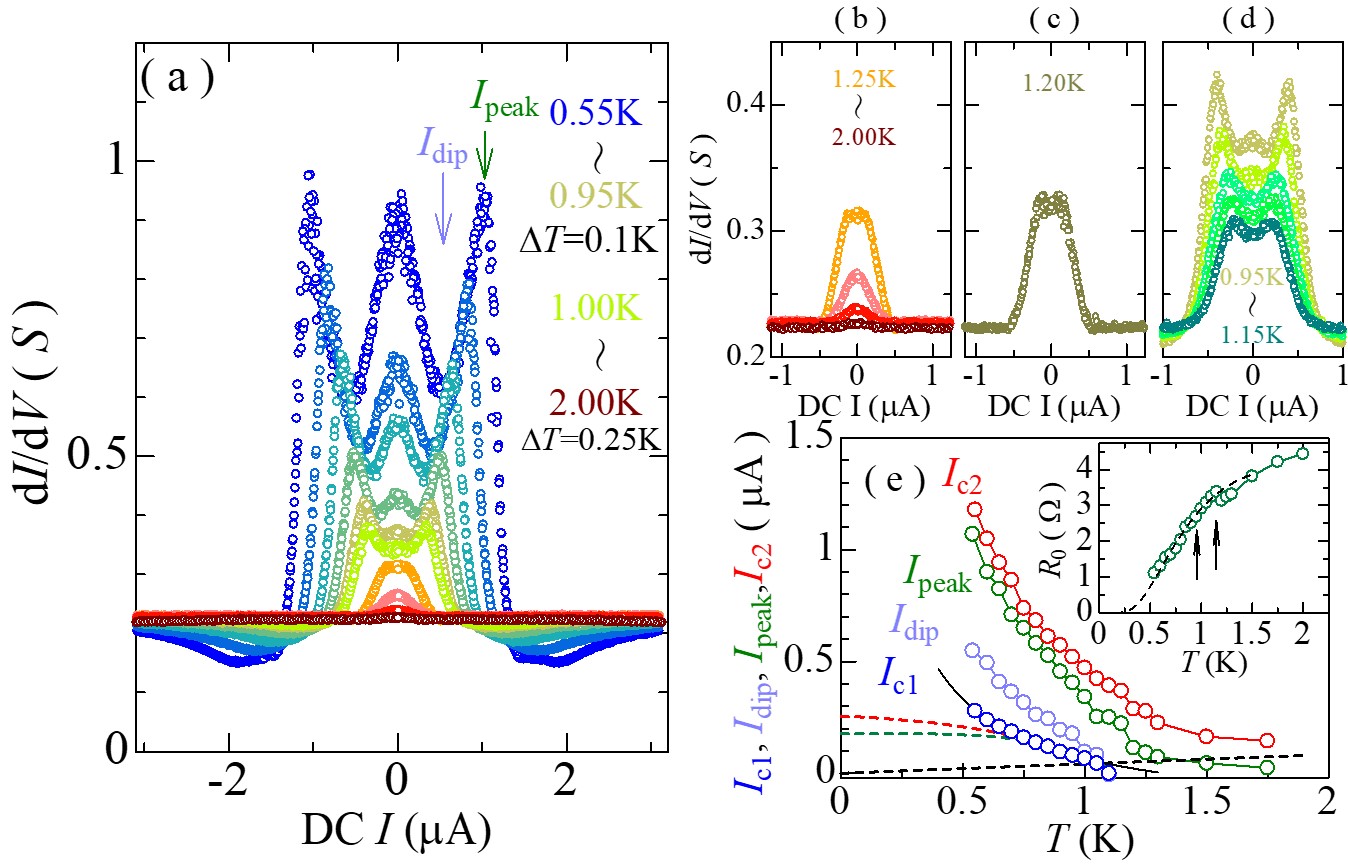}
\caption{\label{fig3}   (Color online) (a) Differential conductance $dI$/$dV$ versus DC-$I$ taken at various temperatures from 0.55 K to 2.00K (0.55 - 0.95 K by 0.1 K steps and 1.00 - 2.00 by 0.25 K steps). Arrows indicate defined characteristic currents $I_{\mathrm{dip}}$ and $I_{\mathrm{peak}}$. (b)-(d) Highlighted temperature evolutions of the spectra. (e) Temperature dependence of $I_{\mathrm{c1}}$, $I_{\mathrm{dip}}$, $I_{\mathrm{peak}}$, and $I_{\mathrm{c2}}$. The solid line indicates the fitting curve of $I_c$ below 1K. The black dashed lines is the competition line of $I_{\mathrm{c}}\hbar /2e = k_{\mathrm{B}}T$. The red and green dashed line indicate approximate analytic results of the general $I_c$-$T$ function with typical parameters for a clean limit and a dirty limit junction, respectively. Inset shows zero-bias temperature dependence of $dV/dI$ ($R_0$). The fitting result using the thermal phase-slip relation was displayed by the dashed line. Two arrows indicate two anomalies of $R_0$.}
\end{figure}

Figure~\ref{fig3}(a) shows the temperature dependence of  $dI/dV$-$I$ of the S/FTI/S junction. Some of the curves exhibiting discernable changes are picked up in Fig.~\ref{fig3}(b)-(d) ($\Delta T =$ 0.25 K and 0.10 K for (b) and (d), respectively). Unique temperature evolution of the induced peak was observed as follows; at first, a typical single peak began to develop at 2 K. At 1.20 K, the peak approached a local maximum value, and the top soon started to split into M-shape. The center of the peak developed as the temperature cooled below 1.00 K. Reflecting this unique temperature evolution, the $T$-dependence of zero-bias resistance ($R_0$) had two anomalies at 1.00 K and 1.20 K as indicated by arrows in the inset of Fig.~\ref{fig3}(e). Most parts of $R_0$ showed a monotonic decrease, and its extrapolation was expected to become zero around 0.2 K. 

The temperature dependence of the critical currents and the characteristic current $I_{\mathrm{dip}}$ ($I_{\mathrm{peak}}$) at the local minimum (maximum) of $dI/dV$-$I$ spectra is summarized in Fig.~\ref{fig3}(e). All curves have a similar trend (i.e., convex downward $T$-dependence). Note that these points do not correspond to the $I_{c1}$ and $I_{c2}$, which can be determined by the local minimum of differential curves of $dI/dV$-$I$ curves.The product $I_{\mathrm{c1}}R_{\mathrm{N,p}}$ ($R_{\mathrm{N,p}} = 3.2 \Omega$) was $\sim$ 0.89 $\mu$eV at 0.55 K, which is also quite small compared with the thermal energy. In contrast, the calculated Josephson coupling energy ($E_{\mathrm{JC1}} = I_{\mathrm{c1}}\hbar /2e = 0.57$ meV) at 0.55 K was significantly larger than the thermal energy. A dashed line in Fig.~\ref{fig3}(e) represents the competition between the thermal energy and Josephson coupling energy $I_{\mathrm{c}}\hbar /2e = k_{\mathrm{B}}T$. The line meets the experimental results around 1.1 K. Therefore, the center of the peak relates to dc-Josephson current. On the other hand, the origin of the side peaks is still unknown and is discussed later.

The $I$-$V$ characteristics with rf irradiation were evaluated to observe $4\pi$-periodic Josephson supercurrent (doubled Shapiro-step). However, finite resistance screened the steps in our measurement. As a result, we observed an almost linear response with a smeared kink structure in the $I$-$V$ curve. Instead, we performed differential conductance $dI/dV$-$V$ measurements under rf irradiation.

\begin{figure}[h]
\centering\includegraphics[width=0.6\tw]{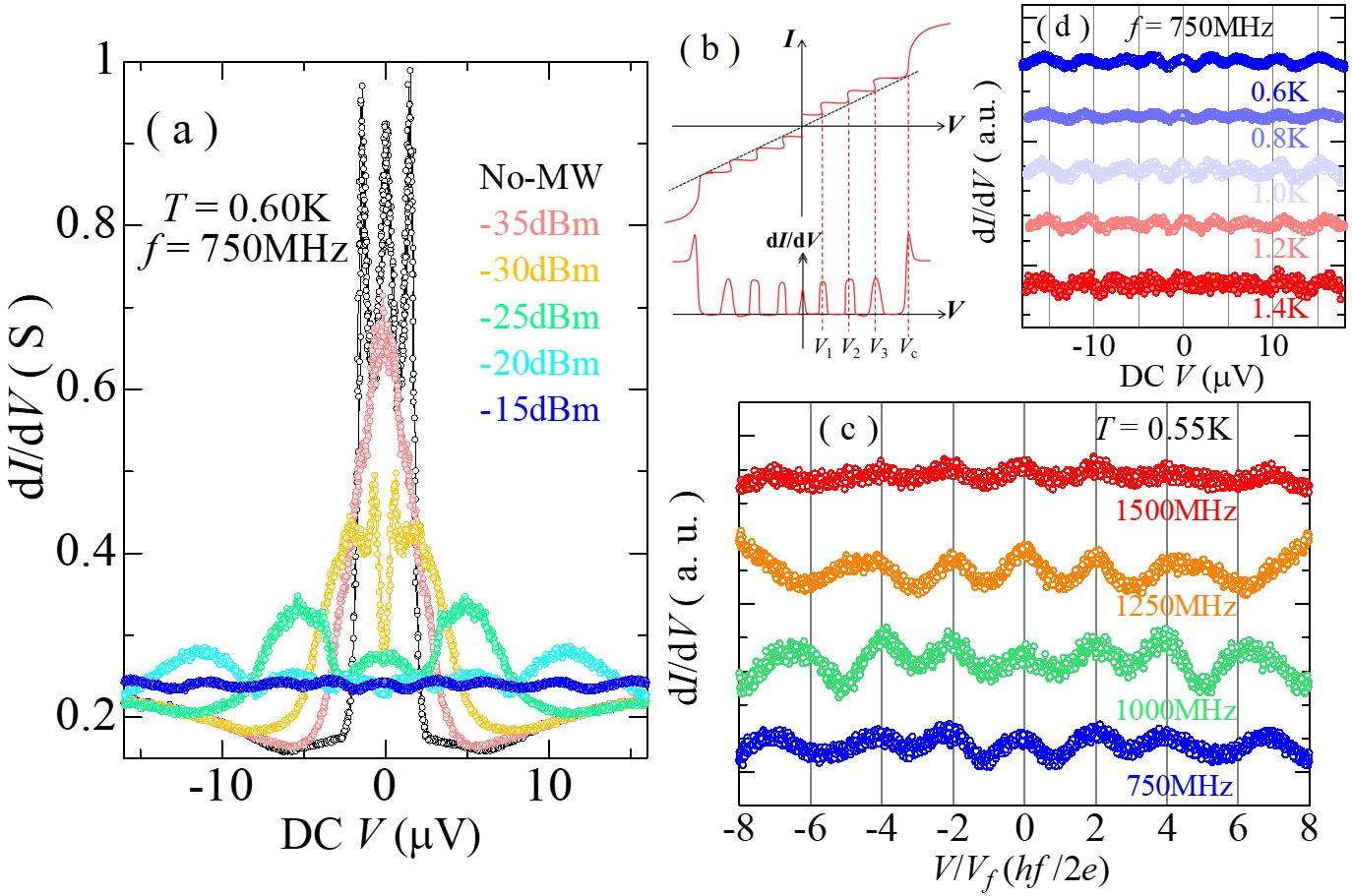}
\caption{\label{fig4}  (Color online) (a) rf-response of induced structure at 0.60 K with frequency of 750 MHz. (b) Expected relation between $I$-$V$ curve and $dI/dV$-$V$ in Shapiro-step response.  (c) Frequency variation of $dI/dV$-$V$ at 0.60 K.  The rf-powers were -15, -10, -4, -4 dBm for 750 to 1500 MHz, respectively. The horizontal axis is normalized by the conventional Shapiro-step voltage $V_f = hf/2e$. (d) Temperature evolutions of $dI/dV$ with rf (750 MHz, -15 dBm) at 0.60 K. }
\end{figure}

The conductance spectra showed a drastic evolution with rf irradiation (Fig.~\ref{fig4}(a)). The peak structure was suppressed rapidly with enhancing the rf power, reflecting the $n$ = 0 Bessel function-like step of the dc-Josephson current. Around -20 dBm, the oscillational components gradually became prominent. Clear oscillational spectra were observed at -15 dBm. In $I$-$V$ curves, the Shapiro-steps usually appear as a step-like signal (upper panel of ~\ref{fig4}(b)) with the step-width voltage of $V_f = hf/2e$, where $h$ and $f$ denote the Planck constant and rf-frequency, respectively.~\cite{ref9} In a $dI/dV$-$V$ picture, the Shapiro-steps look like an oscillating signal (lower panel of Fig.~\ref{fig4}(b)) with the periodicity of $V_f$. The step voltage corresponds to the local maximum of the oscillation. To confirm the origin of the oscillation, rf-frequency and temperature-dependent tests are indispensable.

As a first test, Fig.~\ref{fig4}(c) shows the frequency variation of $dI/dV$-$V$ spectra under moderately high-power rf. The horizontal axis was normalized by the conventional Shapiro-step voltage $V_f$. We used a loop antenna of $\sim$10 mm diameter (the antenna characteristics were focused on $\sim$1 GHz), and thereby the rf power was modulated for each frequency. The oscillational periodicity was proportionally changed with applied rf-frequency within 500 -- 1500 MHz. Second, the $T$-dependence of the oscillatory spectra was evaluated to check the contribution of the resistance. The oscillation periodicity remains unchanged up to 1.4 K (Fig.~\ref{fig4}(d)), while the finite resistance increased by about a factor of four (see inset of Fig.~\ref{fig3}(e)). Therefore, the observed oscillations were irrelevant with the residual resistance, and the above two tests indicated that the signals reflect rf-dependent quantum phenomena including the rf-induced ac-Josephson effect (the Shapiro-steps).

The step widths for all measured frequencies had almost doubled compared to the standard $V_f$ at least up to $n = $ 4 step, which is a hallmark of the $4\pi$-periodicity as expected for the MBS. This result is one of the main results in this paper. The high bias signals showed further high step-widths (approximately $6\pi$-period) above $\sim$ 7 $\mu$V. This trend contrasts with the data for other TI/S junctions and nanowire systems.~\cite{ref10, ref12}

\begin{figure}[h]
\centering\includegraphics[width=0.6\tw]{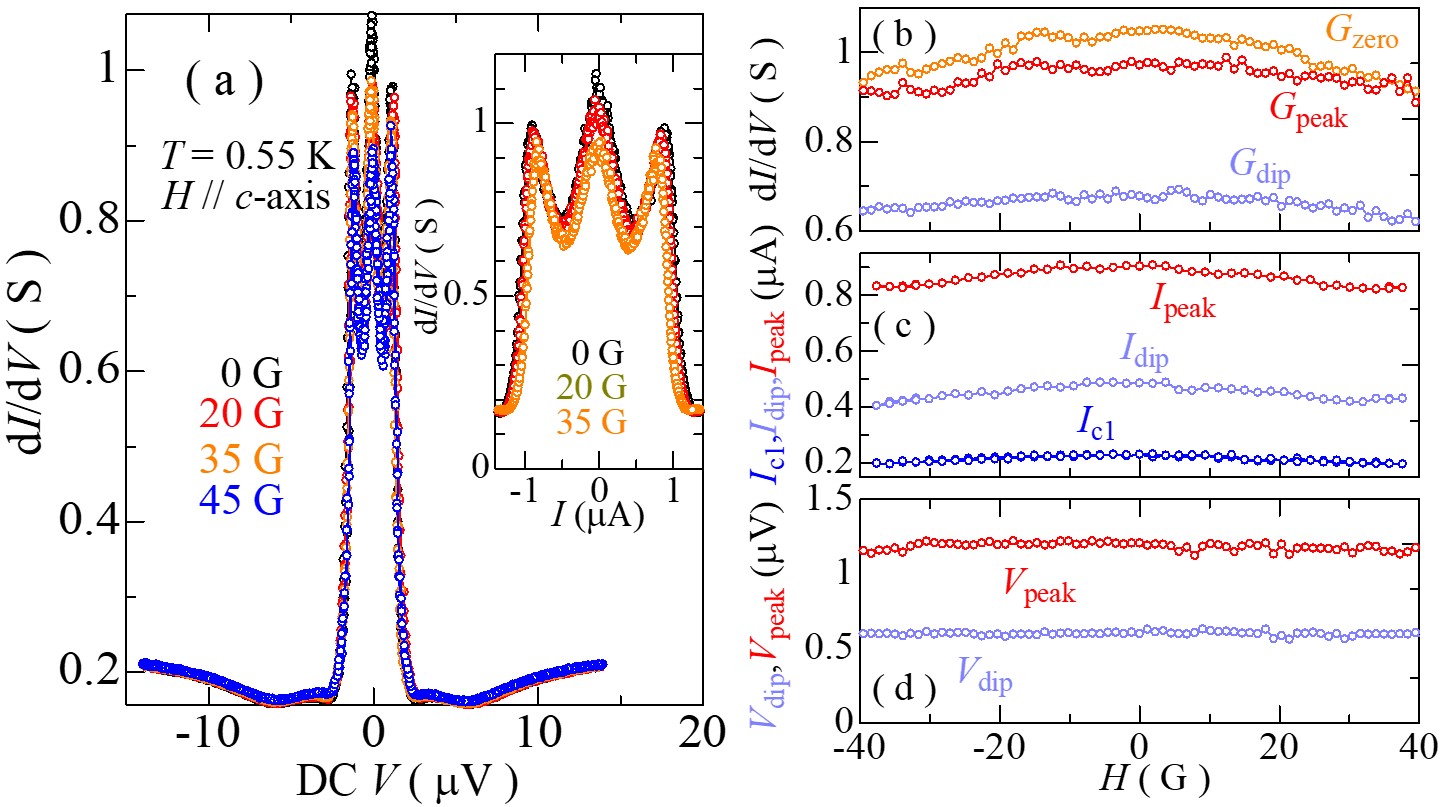}
\caption{\label{fig5}  (Color online) (a) Field dependence of differential conductance spectra at 0.55 K. Inset is enlarged $dI/dV$-$I$ curves at the several magnetic fields. (b)-(d) obtained conductance values, critical currents, and critical voltages defined from Fig.~\ref{fig5}(a). }
\end{figure}

The field dependence of $dI/dV$-$I$ was evaluated to determine the direction of the magnetic moment in the junction (Fig.~\ref{fig5}(a)). The applied magnetic field was perpendicular to the surface ($H//c$-axis). The conductance values at the zero-bias $G_{\mathrm{zero}}$, local maximum $G_{\mathrm{peak}}$, and local minimum $G_{\mathrm{dip}}$ of the trident-shaped peak, as well as the critical currents (voltages) determined at the local maximum/minimum, are shown in Fig.~\ref{fig5}(b)-(d). All characteristics and critical currents decreased with increasing field up to $\pm$40 G, while the  $V_{\mathrm{dip}}$ and the $V_{\mathrm{peak}}$ unchanged.  The $I_{\mathrm{c1}}$-$H$ curve was consistent with the typical Fraunhofer pattern even within the small field range due to the magnetic shield (the estimated field periodicity $\Delta B = \it{\Phi}_{\mathrm{0}}/S$, where $\it{\Phi}_{\mathrm{0}}$ represents the flux quantum and area $S (=100 \mathrm{nm} \times 300 \mathrm{nm})$, is ca. 700 G). If the magnetic moments align along out-of-plane, the $I_{\mathrm{c1}}$-$H$ curve of the $H//c$-axis should be asymmetric because of the hysteresis of the ferromagnetism. Therefore, the present symmetric curve points to an in-plane spin alignment.

\section{Discussions}

In S/N/S junctions, the junction nature and the emergent Cooper pairing change drastically depending on $L/\xi_{\mathrm{N}}$. Thus, we first discuss the general picture of the proximity effect with respect to the distance from electrodes and concluded that the present junction was a weak-link junction which is a promising platform to observe the MBS. Next, we discuss the interpretation of the trident-shaped peak and $4\pi$-periodic Josephson current. Finally, we discuss the possible emergent Cooper pairs and the magnetic doping effect.

The spatial distribution in the proximity effect of an S/FTI/S junction is illustrated as Fig.~\ref{fig1cap}(e), considering Shen et al.~\cite{ref32} In the case of a non-doped S/TI junction, despite the dominant $s$-wave Cooper pairs (ESE) around the interface, spin-triplet Cooper pairs without breaking TRS are expected to dominate as they move away from the interface regions. Meanwhile, in the case of an S/FTI junction, the magnetic moment stabilizes equal spin-triplet pairs and reduces the ESE pairs. As a result, the dominant odd-frequency triplet pairs (OTE) are expected to locate at the red-colored surface region of Fig.~\ref{fig1cap}(e), although details of the induced pairs are sensitive to the doped magnetic moment direction.~\cite{ref31} Such a dominant OTE is one of the essential ingredients to produce the MBS.~\cite{ref27, ref28, ref42} Therefore, the MBS is also located around the red region ($x \sim \xi_{\mathrm{N}}$), similar to the nanowire case,~\cite{ref43} and the present condition $L > 2\xi_{\mathrm{N}}$ is desirable to observe the MBS.

The condition $L > 2\xi_{\mathrm{N}}$ also implies that the induced SC states are weakly connected, and its phase locking is not perfectly developed, leading to non-zero resistance. Generally, the dynamics of a Josephson junction ($\frac{\partial \phi }{\partial t} = \frac{2e}{\hbar }V$) is essentially equal to the Brownian motion of a fictitious particle in a tilted washboard potential on the energy-phase difference ($E$-$\phi$) space. If the potential is well developed, the particle is trapped in one of the metastable potential, leading to the zero resistance (zero voltage). Otherwise, the particle is thermally activated and starts to escape from the trapped potential, resulting in finite resistance. Such a junction is called a phase-slip Josephson junction. The escape rate is governed by temperature, and the temperature dependence of zero-bias resistance (voltage) is known to follow the relation; $R_{\mathrm{0}}$ = $a/T \times $ exp⁡(-$\frac{2E_{\mathrm{J}}}{k_{\mathrm{B}} T}$), where $E_{\mathrm{J}}$ is the Josephson coupling energy and the coefficient a can be written as the product of shunting resistance $R_{\mathrm{in}}$ and $E_{\mathrm{J}}/k_{\mathrm{B}}$ within an overdamped junction condition ~\cite{newref44,newref45}. The excellent fitting result was obtained, as indicated in the inset of Fig. ~\ref{fig3}(e). The obtained parameters ($E_{\mathrm{J}}$ = 0.36 meV and $R_{\mathrm{in}}$ = 4.15 $\Omega$) are comparable with the $E_{\mathrm{JC1}}$ and $R_{\mathrm{N}}$. The fitting line also indicates that low-temperature measurement below ca. 0.2 K is desired to observe the intrinsic nature of the S/N/S junction as future work.

The junction showed two types of structures for different energy scale: the hysteresis for $\sim$1 meV and the peak structure for $\sim$5 $\mu$eV. We considered that the hysteresis displaying the conventional $s$-wave nature does not reflect the junction nature and comes from electrodes (e.g. an interface between yellow-colored larger SC-electrodes and orange narrower ones in Fig.~\ref{fig1cap}(a) or a crystal edge where thin electrodes easily disconnect due to high-step of the thick crystal). The reasons for this consideration are as follow: first, the product of $I_{\mathrm{c}}R_{\mathrm{N}}$ was comparable with that of Nb. Second, the critical temperatures of the hysteretic structure ($\sim$8 K) and the sharp peak ($\sim$1 K) were completely different. Thus, the hysteresis is irrelevant to the proximity effect, and it is the trident peak that reflects the proximity-induced SC states.

One of the main anomalous results was the observation of the trident-shaped peak which had the narrow peak-width less than the thermal energy. The temperature evolution of the peak and the two discontinuities in the $R_0$ indicate the existence of at-least two origins of the peak spectra: the center ZBCP from the dc-Josephson current, and the M-shaped structure coming from other mechanisms. In this paper, we considered two scenarios that explain the origin of the M-shaped structure: a DOS scenario and a triplet components scenario.

In the DOS scenario, the M-shaped conductance and the adjacent dip structures represent a DOS of chiral $p$-wave SC states. Because of the condition $L > 2\xi_{\mathrm{N}}$, the junction may have both the tunneling junction (S/N) and the S/N/S junction nature, and thus potentially show broadened dip or peak structures reflecting quasiparticle DOSs.~\cite{ref3} The observed spectra are almost identical to the calculated spectra of chiral $p$-wave with anisotropic pair amplitude (the parameter $C = 0.3$).~\cite{ref44} The anisotropy is possibly induced by an in-plane magnetization component. However, the present peak-width was quite small compared with the ZBCP in other S/N systems (e.g., nanowire: $\sim 50~\mu$V,~\cite{ref5, ref6} and the Bi-based TI/S junctions: several hundreds of $\mu$V~\cite{ref17, ref19, ref20}). Furthermore, this DOS scenario cannot explain the tiny peak-width less than the thermal energy, which usually conceals such tiny signals without quantum phenomena.

As the other scenario, the M-structure may be derived from quantum tunneling of Cooper pairs (the S/N/S junction nature), which is irrelevant to thermal fluctuation. The peak reflects condensed components of the proximity effect, and the two peaks represent $\uparrow  \uparrow$ and $\downarrow  \downarrow$ states (less $\uparrow  \downarrow$ component), i.e., dominant same spin-triplet Cooper pairing. In this scenario, the peak width should correspond to the energy difference of the two levels, which may relate to any magnetic property. However, detail magnetic properties of $\mu$m-scale thin films are usually different from those of bulk crystals; hence, we have been conducting high-sensitivity magnetic field measurement of microscopic-size crystals using a similar technique to Ref.~\citen{ref45}.

Regarding the dc-Josephson effect, one prominent feature is the low-temperature anomaly of the critical current $I_{\mathrm{c1}}$. It rapidly increased with the cooling process, apparently without saturation. Similar temperature dependence is usually observed in the clean-limit ($\xi_{\mathrm{N}} < \ell_{\mathrm{3D}}$) systems (the red dashed line in the inset of Fig.~\ref{fig3}(e)). However, the present junction is close to be the dirty-limit ($\ell_{\mathrm{3D}} < \xi_{\mathrm{N}}$); i.e., the curves should reach a certain saturated value according to the Kulik-Omel'yanchuk theory (the green dashed line).~\cite{ref46} Furthermore, we did not obtain enough fitting result even using the general formula for the critical current of weak links.~\cite{newref49} Below 1.0 K, our results can be fitted by a function of $I_{\mathrm{c1}} = a/T + b$  ($a =$ 0.268, $b =$ -0.202), as shown by a solid line in Fig.~\ref{fig3}(e). Despite the limited number of measurement points, the fitting range covered $T/T_c \sim $ 0.5-1.0. The low-temperature anomaly obeying $I_{c} \propto T^{-1}$ is theoretically predicted in $p_x$-wave pairs,~\cite{ref47} and another theory predicts convex downward $I_{\mathrm{c}}$-$T$ curves in some unconventional SC junctions.~\cite{ref48} However, the present temperature is not enough for determining a function of $I_{\mathrm{c1}}$-$T$. Though accessing further low-temperature data below $T/T_c < 0.2$ is indispensable, the low-temperature anomaly implies the unconventional nature of the proximity effect. In future, we plan to perform such measurement below 50 mK.

Another anomalous result was the observation of the doubled Shapiro step. The $4\pi$-periodicity is understood as a single e-charge conduction of the Majorana quasiparticles instead of the charge-2e Cooper pairs. Since the MBS is located at the zero energy in the $E$-$\phi$ relation of the ABS of proximity-induced SC states; therefore, the $E$-$\phi$ relationship can intersect the zero energy, forming $4\pi$-periodicity.~\cite{ref9} Namely, Majorana fermions are induced when the phase difference $\phi = \pi$ by rf irradiation.

Frequency and power dependence of the Shapiro steps is critical to distinguish whether the observed $4\pi$-periodicity is the topological origin or not. Some mechanisms, e.g., Landau-Zener process through a highly transparent channel,~\cite{newref52, newref53} derive a transition from conventional Shapiro steps to doubled steps by increasing the frequency and power of the microwave.~\cite{newref54} On the other hand, $4\pi$-periodic contribution induced by the topological origins shows an opposite trend: i.e., the standard steps appear under high-power rf-irradiation.~\cite{ref10, ref12,newref55} The measurement frequency must be less than the inverse of phase relaxation time for the $4\pi$-periodicity:  $f_{\mathrm{rf}} < 1/\tau_{4\pi}$ = $eRI_{\mathrm{c}}^{4\pi}/\hbar$, where $R$ and $I_{\mathrm{c}}^{4\pi}$ are the shunting resistance and the critical current amplitude for the $4\pi$-periodic contribution, respectively, according to the two-channel RSJ model.~\cite{newref56} Thus, low-power and low-frequency experiments are required .~\cite{newref56,newref57} 

In the present junction, almost the same doubled steps were observed within 0.5 - 1.5 GHz. These measurement frequencies are less than the 1/$\tau_{4\pi} \sim$  2 GHz calculated using $R_{in}$ and $I_{c1}$. Regarding the power dependence, it was difficult to follow the step width due to the complicated rf-response, i.e., extra peaks and dips hinder step widths within -30 to -20 dBm. Instead, we plotted the power-dependence of the center peaks, which reflect $I_{c1}$, as shown in Fig. ~\ref{fig6}. The zero-bias $dI/dV$ rapidly decreased by applying the microwave, and then oscillated periodically. The blue line indicates an approximate analytic result using the Bessel function with $n$ = 0. If any transition from $2\pi$- to $4\pi$-periodicity occurs, the periodicity in power dependence is also expected to change. However, such periodic change was not found. This fact implies that the observed $4\pi$-periodic oscillations were not induced by high-power rf irradiation. Therefore, we believe that the observed $4\pi$-periodic Josephson current comes from the topological origin, although we have to check the power dependence of other $n \neq 0$ steps. To confirm this point, we also plan to perform further detail investigation with both low and high power rf-irradiation below 50mK.

\begin{figure}[h]
\centering\includegraphics[width=0.4\tw]{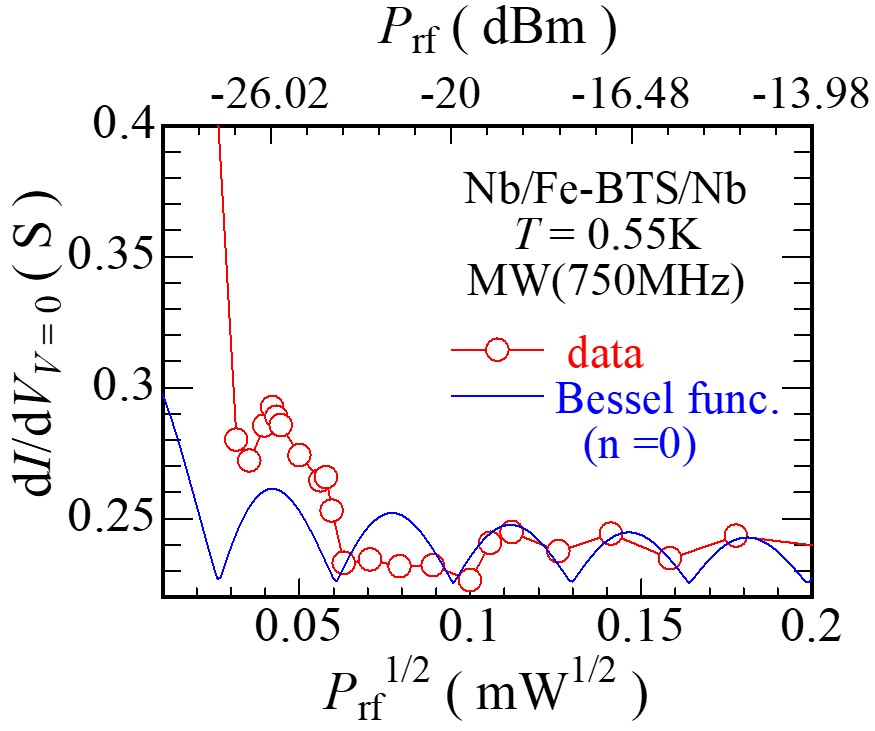}
\caption{\label{fig6}  (Color online) Power dependence of zero-bias $dI/dV$ for the frequency of 750 MHz at 0.55 K. The blue line is the approximate analytic result  of the Bessel function with $n$ = 0. }
\end{figure}

Regarding the origin of the complicated rf-response, several mechanisms are known to produce extra steps or change the step width in $I$-$V$ characteristics, such as the Fiske step~\cite{newref58, newref59} and the fractional-order Shapiro steps.~\cite{newref60} We can distinguish such origins by changing frequency or rf-power, e.g., the Fiske steps appear at the resonant frequencies of cavity modes. These mechanisms, however, cannot explain the present complex response that vanished under high power rf-irradiation. One possible origin that can appear at a limited power range is a chaotic response, which appears in the $I$-$V$ curve as jagged.~\cite{newref61} However, such a chaotic behavior is less likely to emerge in an overdamped situation because nonlinearity is essential for chaotic phenomena.~\cite{newref61} Therefore, an extra investigation is also important to reveal the kinetics of this complex response and to distinguish step widths without a complex hindrance.

Comparing with previous reports, either first-step missing or missing a few steps in the Shapiro-step was found, while higher steps transformed into the standard Shapiro-step pattern.~\cite{ref10, ref12,newref55} A recent report provides one explanation; i.e. thermal poisoning by joule overheating and a high energy rf-power may induce conventional $2\pi$-periodic steps.~\cite{newref62} In contrast, our junction showed further high step widths. One possible reason is the low rf-frequency used in this study, which is several times lower than transition frequency reported other experiments (typically around several GHz~\cite{ref10, ref12}). Since non-doped TI based junctions also showed missing the first step $n$ = $\pm$1 (strained HgTe/Nb ~\cite{newref55} and Bi$_2$Se$_3$/V ~\cite{newref62}) or conventional steps,~\cite{ref51, ref52} the absence of several odd steps and longer period at high bias voltages in our junction suggest that some kinetics against thermal fluctuation may exist in a magnetic TI junction, and such theoretical investigation is required as future work.

Finally, we discuss the doping effect from the aspect of the emergent pairing. The pairing can be intuitively understood by considering a non-doped TI/S case as a first step; i.e., the spin-momentum locking on the Dirac cone suppresses a conventional $s$-wave pairing ($\uparrow \downarrow$--$\downarrow \uparrow$: ESE) due to lack of $\uparrow  \downarrow$ pairing, and as a result, the $p$-wave SC state ($\uparrow \downarrow$+$ \downarrow \uparrow$: ETO) emerges on the interface of TI/S to compensate for the missing $\uparrow  \downarrow$ pairing. In the doped junctions, the pairing depends on the direction of the moments.~\cite{ref31} When the moments are aligned perpendicular to the surface, they lead to the triplet ($\uparrow \uparrow$) $p$-wave pairing (ETO) in the out-of-plane component and the triplet ($\uparrow \downarrow$+$\downarrow \uparrow$) $p$-pairing (ETO) with the TRS on the in-plane component, simultaneously. The in-plane pairing is converted to an odd-frequency triplet ($\uparrow \downarrow$+$\downarrow \uparrow$) $s$-pairing (OTE) at the surface due to lack of translational symmetry. On the other hand, for the in-plane aligned moments (the present case), the spins parallel to the moments form triplet ($\uparrow \uparrow$) pairing (OTE). Therefore, both cases should lead to OTE pairs. Theoretical studies conducted on both cases in similar situations support the emergence of odd pairs on magnetic TI/S junctions.~\cite{ref31, ref52, ref53, ref54} In short, the magnetic doping reduced conventional ESE pairs and stabilized OTE pairing that has the MBS. The doping also leads to the short coherence length ($\sim$40 nm); whereas non-doped TIs of Bi$_{\mathrm{2}}X_{\mathrm{3}}$ ($X =$ Se, Te) have long coherence length ($\sim$1 $\mu$m).~\cite{ref52, ref55} Therefore, magnetic doping played an essential role in the observation of MBS.

The induced Majorana mode here is not locally static in contrast to edge modes in other low-dimensional systems. In the context of braiding operation, a local Majorana mode is also indispensable. We believe that vortex on the S/FTI/S junction also hosts the local MBS, and thus this system potentially exhibits both local and nonlocal Majorana fermions, which provide spatial advantages and widespread options for future braiding operation. Thus, magnetic TI-based junctions are suitable platforms for pursuing Majorana physics and unconventional superconductivity.

\section{Summary}

We examined the transport properties of the S/Fe-BTS/S junction, and experimentally probed the existence of unconventional SC states. The unique phenomena in the junction were the emergence of the exceptional trident-shaped conductance peak, the low-temperature anomaly in $I_{\mathrm{c1}}$-$T$, and the $4\pi$-periodic Josephson effect, which implies the possibility of existing MBS induced by rf irradiation. Those results suggest that the magnetic Josephson junctions are a promising platform for unconventional SC studies. The spatial advantage of electrodes on the 2D surface of 3D TI potentially accelerates the establishment of fundamental technology for future quantum computing and observation of interplay between unconventional superconductors and exotic quasiparticles.

\begin{acknowledgment}
We acknowledge our fruitful discussion with Yukio Tanaka. This study was supported by JST CREST (Grant No. JPMJCR16F2) and KAKENHI (Grant Nos. JP15H05851, 15H05853, 16H03847, and 18H01243). The crystal growth and characterization were supported by the Collaborative Research Projects of Laboratory for Materials and Structures, Institute of Innovative Research, Tokyo Institute of Technology. The fabrication process was performed at the AIST Nano-Processing Facility, supported by "Nanotechnology Platform Program" of the Ministry of Education, Culture, Sports, Science, and Technology (MEXT), Japan, Grant Number JPMXP09F19NM0010.
\end{acknowledgment}


\begin{thebibliography}{99}
\bibitem{ref1} S. Kashiwaya and Y. Tanaka, Rep. Prog. Phys. {\bf 63}, 1641 (2000).
\bibitem{ref2} Y. Asano, Y. Tanaka, M. Sigrist, and S. Kashiwaya, Phys. Rev. B {\bf 71}, 214501 (2005).
\bibitem{ref3} G. E. Blonder, M. Tinkham, and T. M. Klapwijk, Phys. Rev. B {\bf 25}, 4515 (1982).
\bibitem{ref4} M. Octavio, M. Tinkham, G.E. Blonder, and T.M. Klapwijk, Phys. Rev. B {\bf 27}, 6739 (1983).
\bibitem{ref5} V. Mourik, K. Zuo, S. M. Frolov, S. R. Plissard, E. P. A. M. Bakkers, and L. P. Kouwenhoven, Science {\bf 336}, 1003 (2012). 
\bibitem{ref6} M. T. Deng, C. L. Yu, G. Y. Huang, M. Larsson, P. Caroff, and H. Q. Xu, Nano Lett. {\bf 12}, 6414 (2012).
\bibitem{ref7} J Chen, P. Yu, J. Stenger, M. Hocevar, D. Car, S. R. Plissard, E. P. A. M. Bakkers, T. D. Stanescu, S. M. Frolov, Sci. Adv. {\bf 3}, e1701476 (2017).
\bibitem{ref8} F. Nichele, A. C. C. Drachmann, A. M. Whiticar, E. C. T. O'Farrell, H. J. Suominen, A. Fornieri, T. Wang, G. C. Gardner, C. Thomas, A. T. Hatke, P. Krogstrup, M. J. Manfra, K. Flensberg, and C. M. Marcus, Phys. Rev. Lett. {\bf 119}, 136803 (2017).
\bibitem{ref9} A Y. Kitaev,  Phys.-Usp. {\bf 44}, 131 (2001).
\bibitem{ref10} L. P. Rokhinson, X. Liu, and J. K. Furdyna, Nat. Phys. {\bf 8}, 795 (2012).

\bibitem{ref12} E. Bocquillon, R. S. Deacon, J. Wiedenmann, P. Leubner, T. M. Klapwijk, C. Brüne, K. Ishibashi, H. Buhmann and L. W. Molenkamp, Nat. Nanotech. {\bf 12}, 137 (2017).
\bibitem{ref13} E. Majorana, Nuovo Cimento {\bf 14}, 171 (1937).
\bibitem{ref14}  L. Fu and C. L. Kane,  Phys. Rev. B {\bf 79}, 161408(R) (2009).
\bibitem{ref15} L. Fu and C. L. Kane, Phys. Rev. Lett. {\bf 100}, 096407 (2008). 
\bibitem{ref16} J.-P. Xu, M.-X. Wang, Z. L. Liu, J.-F. Ge, X. Yang, C. Liu, Z. A. Xu, D. Guan, C. L. Gao, D. Qian, Y. Liu, Q.-H. Wang, F.-C. Zhang, Q.-K. Xue, and J.-F. Jia, Phys Rev. Lett. {\bf 114}, 017001 (2015).
\bibitem{ref17} F. Yang, Y. Ding, F. Qu, J. Shen, J. Chen, Z. Wei, Z. Ji, G. Liu, J. Fan, C. Yang, T. Xiang, and L. Lu, Phys. Rev. B {\bf 85}, 104508 (2012).
\bibitem{ref18} H. Li, T. Zhou, J. He, H.-W. Wang, H. Zhang, H.-C. Liu, Y. Yi, C. Wu, K. T. Law, H. He, and J. Wang, Phys. Rev. B {\bf 96}, 075107 (2017).
\bibitem{ref19} G. Koren, T. Kirzhner, E. Lahoud, K. B. Chashka, and A. Kanigel, Phys. Rev. B {\bf 84}, 224521 (2011).
\bibitem{ref20} M. P. Stehno, N. W. Hendrickx, M. Snelder, T. Scholten, Y. K Huang, M. S. Golden, and A. Brinkman, Semicond. Sci. Technol. {\bf 32}, 094001 (2017).
\bibitem{ref21} S. Charpentier, L. Galletti, G. Kunakova, R. Arpaia, Y. Song, R. Baghdadi, S. M. Wang, A. Kalaboukhov, E. Olsson, F. Tafuri, D. Golubev, J. Linder, T. Bauch, and F. Lombardi, Nat. Commun. {\bf 8}, 2019 (2017). 
\bibitem{ref22} A. Kitaev, Annals of Physics {\bf 321}, 2 (2006).
\bibitem{ref23} S. R. Elliott and M. Franz, Rev. Mod. Phys. {\bf 87}, 137 (2015).
\bibitem{ref24} Q.-F. Liang, Z. Wang and X. Hu, EPL {\bf 99}, 50004 (2012).
\bibitem{ref25} H. Ebisu, E. Sagi, Y. Tanaka, and Y. Oreg, Phys. Rev. B {\bf 95}, 075111 (2017).
\bibitem{ref26} J. Manousakis, A. Altland, D. Bagrets, R. Egger, and Y. Ando, Phys. Rev. B {\bf 95}, 165424 (2017).
\bibitem{ref27} Y. Tanana, M. Sato, N. Nagaosa, J. Phys. Soc. Jpn. {\bf 81}, 011013 (2012).
\bibitem{ref28} Y. Asano and Yukio Tanaka,  Phys. Rev. B {\bf 87}, 104513 (2013).
\bibitem{ref29} Y. Asano, Yukio Tanaka, and Alexander A. Golubov, Phys. Rev. Lett. {\bf 98}, 107002 (2007).
\bibitem{ref30} R. S. Keizer, S. T. B. Goennenwein, T. M. Klapwijk, G. Miao, G. Xiao, and A. Gupta, Nature (London) {\bf 439}, 825 (2006). 
\bibitem{ref31} P. Burset, B. Lu, G. Tkachov, Y. Tanaka, E. M. Hankiewicz, and B. Trauzettel, Phys. Rev. B {\bf 92}, 205424 (2015).
\bibitem{ref32}  J. Shen, Y. Ding, Y. Pang, F. Yang, F. Qu, Z. Ji, X. Jing, J. Fan, G. Liu, C. Yang, G. Chen, and L. Lu,  arxiv: 1303.5598v3 [cond-mat.mes-hall].
\bibitem{ref33} Y. L. Chen, J.-H. Chu, J. G. Analytis, Z. K. Liu, K. Igarashi, H.-H. Kuo, X. L. Qi, S. K. Mo, R. G. Moore, D. H. Lu, M. Hashimoto, T. Sasagawa, S. C. Zhang, I. R. Fisher, Z. Hussain, Z. X. Shen, Science (sup. mater.) {\bf 329}, 659 (2010).
\bibitem{ref34} J. Xiong, A.C. Petersen, D. Qu, Y.S. Hor, R.J. Cava, N.P. Ong, Physica E {\bf 44}, 917 (2012).
\bibitem{ref35} M. Neupane, S.-Y. Xu, L. A. Wray, A. Petersen, R. Shankar, N. Alidoust, Chang Liu, A. Fedorov, H. Ji, J. M. Allred, Y. S. Hor, T.-R. Chang, H.-T. Jeng, H. Lin, A. Bansil, R. J. Cava, and M. Z. Hasan, Phys. Rev. B {\bf 85}, 235406 (2012).
\bibitem{ref36} K. Miyamoto, A. Kimura, T. Okuda, H. Miyahara, K. Kuroda, H. Namatame, M. Taniguchi, S. V. Eremeev, T. V. Menshchikova, E. V. Chulkov, K. A. Kokh, and O. E. Tereshchenko, Phys. Rev. Lett. {\bf 109}, 166802 (2012).
\bibitem{ref37} Z. Ren, A. A. Taskin, S. Sasaki, K. Segawa, and Y. Ando, Phys. Rev. B {\bf82}, 241306(R) (2010).
\bibitem{ref38} Y. Tanaka and M. Tsukada, Phys. Rev. B {\bf37}, 5087 (1988).
\bibitem{ref39} T. Akazaki, T. Kawakami, and J. Nitta, J. Appl. Phys. {\bf66}, 6121 (1989). 
\bibitem{ref40} M. D. Sherrill and H. H. Edwards, Phys. Rev. Lett. {\bf 6}, 460 (1961).
\bibitem{ref41} V. Ambegaokar and A. Baratoff, Phys. Rev. Lett. {\bf 10}, 486 (1963). 
\bibitem{newref41} R. Yano, Hishiro T. Hirose, Kohei Tsumura, Shuhei Yamamoto, Masao Koyanagi, Manabu Kanou, Hiromi Kashiwaya, Takao Sasagawa, and Satoshi Kashiwaya, Condens. Matter {\bf 4}, 9 (2019).
\bibitem{ref42} M. Snelder, A. A. Golubov, Y. Asano, and A Brinkman, J. Phys.: Condens. Matter {\bf 27}, 315701 (2015)
\bibitem{ref43} D. Chevallier, D. Sticlet, P. Simon, and C. Bena, Phys. Rev. B {\bf 87}, 165414 (2013).
\bibitem{newref44} Y. M. Ivanchenko and L. A. Zil'berman, Sov. Phys. JETP {\bf28}, 1272 (1969).
\bibitem{newref45} V. Ambegaokar and B. I. Halperin, Phys. Rev. Lett.  {\bf22}, 1364 (1969).
\bibitem{ref44} S. Kashiwaya, H. Kashiwaya, K. Saitoh, Y. Mawatari, and Y. Tanaka, Physica E {\bf 55}, 25 (2014).
\bibitem{ref45} Y. Miura, S. Kashiwaya, and S. Nomura, Jpn. J. Appl. Phys. {\bf 56}, 04CK03 (2017).
\bibitem{ref46}  I. O. Kulik and A. N. Omel'yanchuk, Zh. Eksp. Teor. Fiz. Pis. Red. {\bf 21}, 216 (1975); JETP Lett. {\bf 21}, 96 (1975).
\bibitem{newref49} H. Ohta, T. Matsui, Y. Uchikawa, K. Kobayashi, and M. Aono, Physica C {\bf 352}, 186 (2001).
\bibitem{ref47} Y. Asano, Y. Tanaka, and S. Kashiwaya, Phys. Rev. Lett. {\bf 96}, 097007 (2006).
\bibitem{ref48} T. Yokoyama, Y. Tanaka, and A. A. Golubov, Phys. Rev. B {\bf 75}, 094514 (2007).
\bibitem{newref52} J. D. Sau, E. Berg, and B. I. Halperin, arXiv: 1206.4596 [cond-mat.mes-hall].
\bibitem{newref53} L. Landau, Phys. Z. Sowjetunion {\bf 2}, 46 (1932); C. Zener, Proc. R. Soc. London Ser. A {\bf 137}, 696 (1932).
\bibitem{newref54} P. Baars, A. Richter, and U. Merkt,   Phys. Rev. B {\bf 67}, 224501 (2003).
\bibitem{newref55} J. Wiedenmann, E. Bocquillon, R. S. Deacon, S. Hartinger, O. Herrmann, T. M. Klapwijk, L. Maier, C. Ames, C. Br\"{u}ne, C. Gould, A. Oiwa, K. Ishibashi, S. Tarucha, H. Buhmann and L. W. Molenkamp, Nat. Commun. {\bf 7}, 10303 (2016).
\bibitem{newref56} K. L. Calvez, L. Veyrat, F. Gay, P. Plaindoux, C. B. Winkelmann, H. Courtois, and B. Sac\'{e}p\'{e}, Commun. Phys. {\bf 2}, 4 (2019).
\bibitem{newref57} Jay D. Sau and F. Setiawan, Phys. Rev. B {\bf 95}, 060501(R) (2017).
\bibitem{newref58} D.D.  Coon  and  M.D.  Fiske, Phys.  Rev. {\bf 138},  A744 (1965).
\bibitem{newref59} J.-J. Chang, Phys. Rev. B {\bf 28}, 1276 (1983).
\bibitem{newref60} R. C. Dinsmore III, M.-H. Bae, and A. Bezryadin, Appl. Phys. Lett. {\bf 93}, 192505 (2008).
\bibitem{newref61} R. L. Kautz, Rep. Prog. Phys. {\bf 59}, 935 (1996).
\bibitem{newref62} K. L. Calvez et al., Commun. Physics {\bf 2}, 4 (2019).
\bibitem{ref51} L. Galletti, S. Charpentier, M. Iavarone, P. Lucignano, D. Massarotti, R. Arpaia, Y. Suzuki, K. Kadowaki, T. Bauch, A. Tagliacozzo, F. Tafuri, and F. Lombardi, Phys. Rev. B {\bf 89}, 134512 (2014). 
\bibitem{ref52} M. Veldhorst, M. Snelder, M. Hoek, T. Gang, V. K. Guduru, X. L. Wang, U. Zeitler, W. G. van der Wiel, A. A. Golubov, H. Hilgenkamp and A. Brinkman, Nat. Mater. {\bf 11}, 417 (2012).
\bibitem{ref53} Y. Tanaka, T. Yokoyama, and N. Nagaosa, Phys. Rev. Lett. {\bf 103}, 107002 (2009).
\bibitem{ref54} T. Yokoyama, Phys. Rev. B {\bf 86}, 075410 (2012)
\bibitem{ref55} F. Yang, F. Qu, J. Shen, Y. Ding, J. Chen, Z. Ji, G. Liu, J. Fan, C. Yang, L. Fu, and L. Lu, Phys. Rev. B {\bf 86}, 134504 (2012).

 



\end{thebibliography}
\end{document}